\documentstyle[aps,multicol,epsf]{revtex} 

\makeatletter
\def\lsim{\mathrel{\mathpalette\gl@align<}}
\def\gsim{\mathrel{\mathpalette\gl@align>}}
\def\gl@align#1#2{\lower.6ex\vbox{\baselineskip\z@skip\lineskip\z@
    \ialign{$\m@th#1\hfil##\hfil$\crcr#2\crcr\sim\crcr}}}
\makeatother

\begin{document}

\bibliographystyle{prsty}
\draft	
\title{Two-Impurity Kondo Effect in Double-Quantum-Dot Systems\\
--- Effect of Interdot Kinetic Exchange Coupling ---}
\author{Wataru Izumida\cite{email}}
\address{Department of Applied Physics, Hokkaido
University, Sapporo 060-8628, Japan}
\author{Osamu Sakai} 
\address{Department of Physics, Tokyo Metropolitan
University, Tokyo 192-0397, Japan}
\maketitle
\begin{abstract}
Tunneling conductance through two quantum dots,
which are connected in series to left and right
leads, is calculated by using the numerical
renormalization group method.
As the hopping between the dots increases from
very small value, the following states
continuously appear; (i) Kondo singlet state of
each dot with its adjacent-site lead, (ii) singlet
state between the local spins on the dots, and
(iii) double occupancy in the bonding orbital of
the two dots.
The conductance shows peaks at the transition
regions between these states.
Especially, the peak at the boundary between (i)
and (ii) has the unitarity limit value of
$2e^{2}/h$ because of coherent connection through
the lead-dot-dot-lead.
For the strongly correlated cases, the
characteristic energy scale of the coherent peak
shows anomalous decrease relating to the quantum
critical transition known for the two-impurity
Kondo effect.
The two dots systems give the new realization of
the two-impurity Kondo problem.
\end{abstract}

\begin{multicols}{2}
\narrowtext

\section{Introduction}
\label{sec:1}

Dilute magnetic impurities in metal bring the
single-impurity Kondo effect
\cite{rf:Hewson,rf:Yosida}.
The anti-ferromagnetic coupling between spins on
impurities $J$, such as the RKKY interaction,
would compete with the Kondo effect.
To study such competition effect, the two
impurities in metal have been studied extensively
\cite{rf:Jones8701,rf:Jones8801,rf:Jones8901,rf:Jones9101,rf:Jones8902,rf:Sakai_2imp9001,rf:Sakai_2imp920102,rf:Fye8701,rf:Fye8901,rf:Fye9401,rf:Affleck9201,rf:Affleck9501}.
If the Kondo binding energy is much larger than
$J$ ($T_{\rm K} \gg J$), each local spin on the
magnetic impurity forms the Kondo singlet state
with the conduction electrons.
On the other hand for $T_{\rm K} \ll J$, the two
local spins form the local spin singlet state.
From the numerical renormalization group (NRG)
calculation, Jones {\it et al.} had pointed out
that the transition between the two states occurs
as a quantum critical phenomenon
\cite{rf:Jones8801}.
However the advanced investigation made clear that
the critical transition is an artifice of the
model neglecting the parity splitting terms, such
as the d-d hopping term between the impurity atoms
\cite{rf:Sakai_2imp9001}.
In this paper we will investigate the effect of
the d-d hopping term in double-quantum-dot (DQD)
systems in detail, and give the new realization to
the two-impurity Kondo problem.

It might be difficult to observe the two-impurity
effect in metal systems as pointed out by previous
studies, because the alloy contains many types of
impurity pairs, and because the coupling between
impurities is fixed in each material.
Recently, the Kondo effect is observed in the
single quantum dot systems
\cite{rf:DGG9801,rf:Cronenwett9801,DGG9802,rf:Schmid9801,rf:Simmel9901}.
The experimental data show good agreement with the
results of numerical calculation based on the
single-impurity Anderson model
\cite{rf:Izumida.SGL}.
These works demonstrated that the quantum dot
systems are suitable for sensitive experiment of
the Kondo problem.
On the DQD systems, each dot corresponds to an
impurity atom, and the coupling between the dots
can be changed freely by applying the split gate
voltage between the dots \cite{rf:EXP_2DOT}.
It would be expected that we can investigate the
two-impurity effect systematically in the DQD
systems.

For the DQD systems, which the two dots are
connected to the left lead and the right lead in
series as `lead-dot-dot-lead', there are several
theoretical works including the Kondo effect
\cite{rf:Ivanov9701,rf:Pohjola9701,rf:Aono9801,rf:Izumida9901,rf:Georges9901,rf:Izumida.LT22}.
We have reported the large enhancement of the
tunneling conductance through the two dots when
the condition $J_{\rm LR}^{\rm eff} \sim T_{\rm
K}^{0}$ holds, by using the NRG calculation
\cite{rf:Izumida9901}.
Here $J_{\rm LR}^{\rm eff}=4 t^{2}/U$ is the
anti-ferromagnetic kinetic exchange coupling
between the two dots, $t$ is the hopping between
the two dots, $U$ is the Coulomb repulsion on the
dot, and $T_{\rm K}^{0}$ is the Kondo temperature
at $t=0$.
We note that the anti-ferromagnetic coupling is
the inevitable effect due to the kinetic process
$t$ and the Coulomb repulsion on the dot $U$.
There are investigations with the slave boson mean
field theory (SBMFT).
Aono {\it et al.} had already studied the same
model of us, however they could not find the
relation $J_{\rm LR}^{\rm eff} \sim T_{\rm K}^{0}$
on the peak of the conductance pointed out by us
because the SBMFT can not treat the kinetic
exchange process properly \cite{rf:Aono9801}.
Georges {\it et al.} introduced the
anti-ferromagnetic coupling $J$ between the two
dots by artifice in the model, and discussed the
effect related to the critical transition on the
conductance by using the SBMFT
\cite{rf:Georges9901}.
However the introduction of the artificial $J$ in
the model and the calculation of the conductance
within the SBMFT framework bring some questions as
follows:
(a) Does the effect of anti-ferromagnetic coupling
pointed by Georges {\it et al.} actually appear in
the DQD systems?, because the hopping itself
breaks the quantum critical transition
\cite{rf:Sakai_2imp9001,rf:Sakai_2imp920102}.
If it appears, however, (b) how the conflicting
effects of $t$, the kinetic exchange coupling that
would cause the critical transition and the parity
splitting that suppress the critical transition,
compete?
And then, (c) how they appear in the conductance?
Since the SBMFT could not treat the kinetic
exchange process properly, this approximation for
the two-impurity Kondo problem like the DQD
systems seems to be unfavorable.
The reliable calculation is necessary for such a
sensitive problem.

In this paper, we present the detailed
investigation of the Kondo effect in the DQD
systems.
The numerical calculation is performed by using
the NRG method.
This numerical method is known to be a reliable
one for the two-impurity Kondo problem
\cite{rf:Jones8701,rf:Jones8801,rf:Jones8901,rf:Jones9101,rf:Sakai_2imp9001,rf:Sakai_2imp920102,rf:Affleck9501}.
We calculate the tunneling conductance through the
two dots.
We note that some preliminary results were
presented at SCES98 \cite{rf:Izumida9901}, and the
one of the central results was presented at LT22
\cite{rf:Izumida.LT22}.

We find that the following states continuously
appear when the hopping between the two dots
increases from very small value;
(i) Kondo singlet state ($t \ll U$, $J_{\rm
LR}^{\rm eff} \ll T_{\rm K}^{0}$),
(ii) singlet state between local spins on the dots
($t \ll U$, $J_{\rm LR}^{\rm eff} \gg T_{\rm
K}^{0}$),
and (iii) double occupancy in the bonding orbital
of the two dots ($t \gsim U$).
The conductance shows peaks at the transition
regions between these states.
The `main peak' at the boundary between (i) and
(ii) with the condition $J_{\rm LR}^{\rm eff} \sim
T_{\rm K}^{0}$ has the unitarity limit value of
$2e^{2}/h$ because of coherent connection through
the lead-dot-dot-lead.
Especially for the strongly correlated cases, the
width of the main peak becomes very narrow and the
characteristic temperature of the peak is largely
suppressed compared with the Kondo temperature of
the single dot systems $T_{\rm K}^{0}$.
These anomalies of the main peak closely relate to
the quantum critical phenomenon in the
two-impurity Kondo problem.
The quantitative calculation in this paper gives
the new realization for the two-impurity Kondo
problem, and suggests the possibility of the
systematic study of the anomalous two-impurity
Kondo effect in the DQD systems.

The formulation is presented in \S \ref{sec:2}.
The numerical results are presented in \S
\ref{sec:3}.
The summary and discussion are given in \S
\ref{sec:4}.

\section{Formulation}
\label{sec:2}

We investigate the following model Hamiltonian for
the DQD systems that the two dots are connected to
the left lead and the right lead in series;
\begin{eqnarray}
   H & = & H_{\rm l} + H_{\rm d} + H_{\rm l-d},
   \label{eq:total_H_TD}\\
   H_{\rm l} & = & 
                     \sum_{k \sigma} 
                     \varepsilon_{k} 
                     c_{{\rm L} k \sigma}^{\dagger} c_{{\rm L} k \sigma}
                 +
                     \sum_{q \sigma} 
                     \varepsilon_{q} 
                     c_{{\rm R} q \sigma}^{\dagger} c_{{\rm R} q \sigma},
   \label{eq:H_l_TD}\\
   H_{\rm d} & = & 
                     \varepsilon_{\rm d, L}
                     \sum_{\sigma}
                     n_{{\rm d, L} \sigma}
                     + 
                     \varepsilon_{\rm d, R}
                     \sum_{\sigma}
                     n_{{\rm d, R} \sigma} \nonumber\\
             & &     + (-t \sum_{\sigma} d_{{\rm L} \sigma}^{\dagger} d_{{\rm R} \sigma} + {\rm h.c.}) \nonumber \\
             & &     + 
                     U_{\rm L} n_{{\rm d, L} \uparrow} n_{{\rm d, L} \downarrow}
		     +
                     U_{\rm R} n_{{\rm d, R} \uparrow} n_{{\rm d, R} \downarrow},
   \label{eq:H_d_TD}\\
   H_{\rm l-d} & = & \sum_{k \sigma}
                     V_{{\rm L} k} 
                     d_{{\rm L} \sigma}^{\dagger} c_{{\rm L} k \sigma}
                     +
                     \sum_{q \sigma}
                     V_{{\rm R} q} 
                     d_{{\rm R} \sigma}^{\dagger} c_{{\rm R} q \sigma} 
                     + {\rm h.c.}
   \label{eq:H_l-d_TD}
\end{eqnarray}
$H_{\rm l}$ gives the electrons in the left
and right leads.
$H_{\rm d}$ gives that in the left and right dots.
$H_{\rm l-d}$ gives the tunneling between the left
lead and the left dot, and between the right lead
and the right dot.
The suffices ${\rm L}$, ${\rm R}$ mean the left
and the right, respectively.
$c_{{\rm L} k \sigma}$ is the annihilation
operator of the electron in the left lead,
$d_{{\rm L} \sigma}$ is that in the left dot.
$n_{\rm d, L \sigma}= d_{{\rm L} \sigma}^{\dagger}
d_{{\rm L} \sigma}$ is the number operator of the
left dot.
$\varepsilon_{k}$ is the energy of the state $k$
in the left lead.
$\varepsilon_{\rm d, L}$ is the energy of the
orbital in the left dot.
The quantity $t$ is the matrix element between the
left and right dots, and we call it as the 'hopping'
between the dots hereafter.
$U_{\rm L}$ is the Coulomb interaction between the
electrons in the left dot.
$V_{{\rm L}}$ is the matrix element between the
left dot and the left lead.

Here we consider only the single orbital in each
of the dots.
This situation is justified in the case that the
typical energy splitting between the orbitals in
the dot is larger than the typical broadening of
the energy levels, $\delta \varepsilon_{\rm d} \gg
\Delta$, and when the temperature is smaller than
the typical Coulomb repulsion between the
electrons in the dots, $T \ll U$
\cite{rf:Izumida.SGL,rf:Izumida9801}.
(The Kondo effect is not important in the case of
$T \gsim U$.)
We consider only the on-site Coulomb interaction
between the electrons.
Furthermore, we consider only the nearest
neighboring tunneling, between the dot and its
adjacent-site lead, between the two dots.
The energies $\varepsilon_{\rm d, L}$,
$\varepsilon_{\rm d, R}$ can be changed by
applying the gate voltage on the dots.
$V_{{\rm L} k}$ ($V_{{\rm R} q}$) can also be
changed by applying the split gate voltage between
the left (right) dot and the left (right) lead.
$t$ can be changed by applying the split gate
voltage between the left dot and the right dot.

In this paper we consider only the symmetric case
with respect to the exchange of the left and the
right.
This situation is written with the following
relations; $\varepsilon_{\rm d} \equiv
\varepsilon_{\rm d, L} = \varepsilon_{\rm d, R}$,
$U \equiv U_{\rm L} = U_{\rm R}$, and $\Delta
\equiv \Delta_{\rm L} = \Delta_{\rm R} = \pi
|V|^{2} \rho_{\rm c}$.
($\Delta$ is the hybridization strength between
the dots and the leads, $V \equiv V_{{\rm L} k } =
V_{{\rm R} k}$, $\rho_{\rm c}$ is the density of
states in the leads.
Here we consider that there are no $k$-dependence
in the matrix element and the density of states.)
The model can be mapped into the two-channel
Anderson Hamiltonian by the unitary transform for
the operators of the dots and the operators of the
leads \cite{rf:Izumida9901}.
Furthermore we consider the situation that there
is one electron in each dot by adjusting the gate
voltage on the dot, $\langle n_{\rm d, L} \rangle
= 1$, $\langle n_{\rm d, R} \rangle = 1$.

We solve the Hamiltonian by using the NRG method,
and calculate the conductance from the current
correlation function within the linear response
theory
\cite{rf:Izumida9901,rf:Izumida9801,rf:Izumida9701}.
(For detailed calculation of the conductance, see
appendix of Ref. \cite{rf:Izumida9701}.)

At zero temperature, the conductance can be
re-written by using the effective parameters of
the fixed point non-interacting Anderson
Hamiltonian as follows;
\begin{eqnarray}
   G & = & \frac{2e^{2}}{h} 
           |\Delta {\cal G}_{\rm e}(0^{+}) - \Delta {\cal G}_{\rm o}(0^{+})|^{2}
	   \nonumber\\
     & = & \frac{2e^{2}}{h} 
           \frac{4 (t^{\rm eff}/\Delta^{\rm eff})^{2} }
                {(1 + (t^{\rm eff}/\Delta^{\rm eff})^{2} )^{2}}.
   \label{eq:G_T=0}
\end{eqnarray}
We have used the relation,
${\cal G}_{p}=z_{p}/(\omega - \varepsilon_{p}^{\rm
eff} + i \Delta_{p}^{\rm eff})$,
$z_{p}=\Delta_{p}^{\rm eff} / \Delta$,
($p = {\rm e, o}$) at $T=0$.
Here the suffix $p$ denote the even and odd parity
orbitals in the two dots.
We note that the even orbital is the bonding
orbital, and the odd orbital is the anti-bonding
orbital.
We now consider the case of $\langle n_{\rm e}
\rangle + \langle n_{\rm o} \rangle = 2$,
then 
$t^{\rm eff} \equiv -\varepsilon_{\rm e}^{\rm eff}
= \varepsilon_{\rm o}^{\rm eff} \ge 0$,
$\Delta^{\rm eff} \equiv \Delta_{\rm e}^{\rm eff}
= \Delta_{\rm o}^{\rm eff}$.
Here $t^{\rm eff}$ is the effective hopping
between the dots, and $\Delta^{\rm eff}$ is the
effective hybridization strength between the leads
and the dots.
At $T=0$, we calculate the effective parameters
from the analysis of the flow chart of the
renormalized energy level structure in the NRG
calculation, and then calculate the conductance
from eq. (\ref{eq:G_T=0}).

\section{Numerical Results}
\label{sec:3}

In numerical calculation we choose the half of the
band width as an energy unit.
The Coulomb repulsion is fixed at $U=0.1$ through
this paper.
We calculate the conductance as a function of the
hopping $t$ for various hybridization strength
$\Delta$.
(As noted in previously, $t$ and $\Delta$ can be
changed by applying the split gate voltage between
the dots, between the dots and the leads,
respectively.)
The gate voltage on the dots is fixed at
$\varepsilon_{\rm d}=-U/2$, then the DQD is in the
half-filled case, i. e. each dot contains one
electron.

In \S \ref{sec:3.1} and \S \ref{sec:3.2} we
present the numerical results at zero temperature
$T=0$, and in \S \ref{sec:3.3} we present the
results at finite temperatures.

\subsection{Conductance in the strongly correlated case}
\label{sec:3.1}

First we present the conductance in the strongly
correlated case with the hybridization strength
satisfying $\Delta/\pi=1.5 \times 10^{-3}$,
($\Delta/\pi U=1.5 \times 10^{-2}$, i.e. $u \equiv
U/\pi \Delta \simeq 6.8$).

\begin{figure}[hbtp]
   \centerline{\epsfxsize=8.5cm\epsfbox{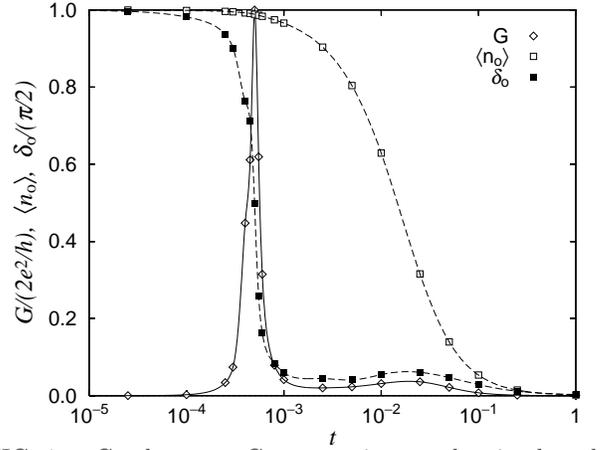}}
   \caption
   {
Conductance $G$, occupation number in the odd
orbital of the two dots $\langle n_{\rm o}
\rangle$, and phase shift of the odd channel
$\delta_{\rm o}$, as a function of the hopping $t$
at $T=0$.
We note the relations between the even and odd
orbitals, $\langle n_{\rm e} \rangle = 2 -
\langle n_{\rm o} \rangle$, $\delta_{\rm e} =
\pi - \delta_{\rm o}$.
$\Delta / \pi U = 1.5 \times 10^{-2}$.
   } 
\label{fig:01}
\end{figure}

We show the conductance at $T=0$ as a function of
the hopping $t$ in
Fig. \ref{fig:01}.
(The occupation number and the phase shift are
also shown in Fig. \ref{fig:01}.)
There are two peaks in the conductance, the large
peak near $t \sim 5 \times 10^{-4}$, and the small
peak near $t \sim 2 \times 10^{-2}$.
(Hereafter we call the large peak as the `main
peak'.)
Why these peaks appear?
In the later paragraph we will analyze various
quantities for the parameter cases showing the
peaks.
%
%

\begin{figure}[hbtp]
   \centerline{\epsfxsize=8.5cm\epsfbox{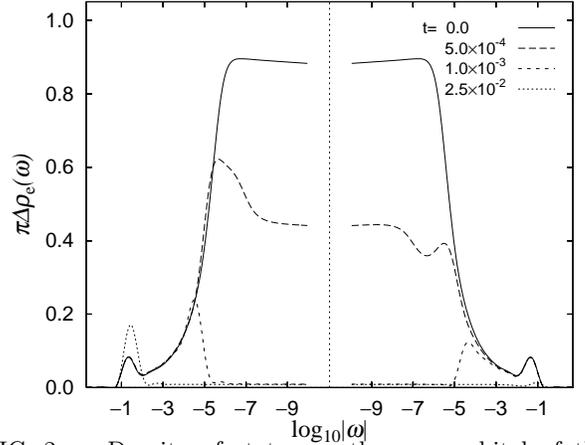}}
   \caption
   {
Density of states on the even orbital of the two
dots, $\rho_{\rm e}(\omega)$.
We note the relation $\rho_{\rm o}(\omega) =
\rho_{\rm e}(-\omega)$, where $\rho_{\rm o}(\omega)$ is
the density of states on the odd orbital.
$\Delta / \pi U =1.5 \times 10^{-2}$.
   }
   \label{fig:02}
\end{figure}
The density of states on the even orbital of the
two dots, $\rho_{\rm e}(\omega)$, for several $t$
cases is shown in
Fig. \ref{fig:02}.
(The relation $\rho_{\rm e}(\omega) = \rho_{\rm
o}(-\omega)$ holds for $\langle n_{\rm e} \rangle
+ \langle n_{\rm o} \rangle=2$, where $\rho_{\rm
o}(\omega)$ is the density of states on the odd
orbital of the two dots.)
At $t=0$, there is the Kondo peak on the Fermi
energy.
(Fermi energy corresponds to $\omega = 0$.)
Naturally, this Kondo peak is caused by the Kondo
singlet states between the left lead and the left
dot, and between the right lead and the right dot.
As $t$ increases to $t = 5 \times 10^{-4}$, the
conductance has the main peak and the strength of
$\rho_{\rm e}(\omega \sim 0)$ becomes half of that
at $t=0$.
In this region we can consider that the Kondo
effect with the spins on the orbitals extending
the two dots, the even and the odd orbitals,
occurs.
As $t$ still increases, the conductance decreases
rapidly, and the strength of $\rho_{\rm e}(\omega
\sim 0)$ is largely suppressed as shown at $t=1.0
\times 10^{-3}$.
This suppression means the disappearance of the
Kondo coupling between the leads and the dots.
$t$ becomes still large, the conductance has the
small peak at $t \sim 2 \times 10^{-2}$.
At $t=2.5 \times 10^{-2}$, the density of states
on the even and odd orbitals have peaks at $\mp
\omega \sim 10^{-1}$, respectively, from
Fig. \ref{fig:02}.
At the same time the occupation numbers begin to
change as shown in
Fig. \ref{fig:01}.
($\langle n_{\rm e} \rangle \simeq 1.5$, $\langle
n_{\rm o} \rangle \simeq 0.5$ at $t \simeq 1.5
\times 10^{-2}$.)

Here we note the following two points:
First, the condition $J_{\rm LR}^{\rm eff} \sim
T_{\rm K}^{0}$ holds at $t \simeq 5 \times
10^{-4}$, where $J_{\rm LR}^{\rm eff} \equiv 4
t^{2}/U$.
($T_{\rm K}^{0} = 3.78 \times 10^{-6}$ is the
Kondo temperature at $t=0$, with the expression
$T_{\rm K}^{0}=\sqrt{U \Delta/2} \exp [-\pi U
/8\Delta + \Delta/2U ]$ \cite{rf:Hewson}, then
$J_{\rm LR}^{\rm eff}/T_{\rm K}^{0} \simeq 2.65$
at $t = 5.0 \times 10^{-4}$.)
Second, as seen from
Fig. \ref{fig:01}, the occupation
numbers of the even and the odd orbitals for $t
\lsim 1.0 \times 10^{-3}$ are almost same with
each other, $\langle n_{\rm e} \rangle \simeq
\langle n_{\rm o} \rangle \simeq 1$.
For $t \gsim 1.0 \times 10^{-1}$, the two
electrons occupy the even orbital.
The border between them is at $t \sim U/4 (=2.5
\times 10^{-2})$.

Above analysis implies the following scenario.
In the case of $t \ll U/4 (= 2.5 \times 10^{-2})$,
the hopping $t$ causes the anti-ferromagnetic
kinetic exchange coupling, $J_{\rm LR}^{\rm eff}$.
For smaller hopping case with $J_{\rm LR}^{\rm
eff} \ll T_{\rm K}^{0}$, there are the Kondo
singlet states between the left lead and the left
dot, and between the right lead and the right dot
each other.
As $t$ increases and then $J_{\rm LR}^{\rm eff}
\gg T_{\rm K}^{0}$, the two local spins on each
dot form the local singlet state.
At the transition region between two states we
have a main peak with the unitarity limit value of
$2e^{2}/h$ in the conductance.
This will indicate that the leads and the dots are
coherently connected by the even and the odd
orbital states.
When $t$ becomes still large and the condition $t
\gsim U/4$ holds, the local spins do not appear,
instead, the two electrons occupy the even
orbital.
The small peak of the conductance reflects the
transition of the electronic states in the DQD.

\begin{figure}[hbtp]
   \centerline{\epsfxsize=8.5cm\epsfbox{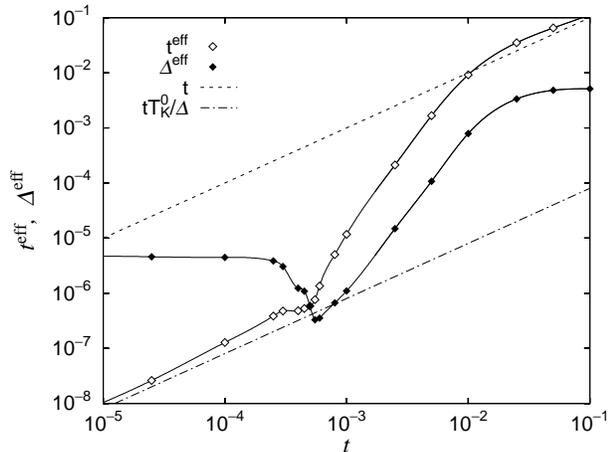}}
   \caption
   {
Effective parameters $t^{\rm eff}$, $\Delta^{\rm
eff}$ of the fixed point non-interacting Anderson
Hamiltonian, given by the analysis of the flow
chart of the renormalized energy level structure
in the NRG calculation.
$\Delta / \pi U = 1.5 \times 10^{-2}$.
   }
   \label{fig:03}
\end{figure}
Here we show the effective parameters $t^{\rm
eff}$ and $\Delta^{\rm eff}$ as a function of $t$
in Fig.
\ref{fig:03}.
(We note that the effective parameters have been
used already for the calculation of the
conductance shown in
Fig. \ref{fig:01}.)
In $t \lsim 10^{-4}$ case, the effective
parameters behave as $t^{\rm eff} \sim t T_{\rm
K}^{0}/\Delta$, $\Delta^{\rm eff} \sim T_{\rm
K}^{0}$.
Then the conductance coincides with the
non-interacting one when one substitutes the
effective parameters into eq. (\ref{eq:G_T=0}).
As $t$ increases with the condition near $J_{\rm
LR}^{\rm eff} \sim T_{\rm K}^{0}$, $\Delta^{\rm
eff}$ once decreases, and it has local minimum, and
then it increases.
At the same time the slope of $t^{\rm eff}$ once
decreases and then increases.
When $t^{\rm eff}$ and $\Delta^{\rm eff}$ coincide
with each other, the conductance has a peak at $t
\sim 5 \times 10^{-4}$, i. e. $J_{\rm LR}^{\rm
eff} \sim T_{\rm K}^{0}$.
As $t$ increases slightly beyond this point, the
conductance sharply decreases because $\Delta^{\rm
eff}$ decreases to the minimum even though $t^{\rm
eff}$ increases.
Here we stress that the relation $J_{\rm LR}^{\rm
eff} \sim T_{\rm K}^{0}$ holds when $t^{\rm eff}
\sim \Delta^{\rm eff}$, and at the same time
$\Delta^{\rm eff}$ becomes very small in the
transition region.
When $t$ increases further, the ratio $t^{\rm
eff}/\Delta^{\rm eff}$ increases gradually in the
region $t \lsim U/4 (= 2.5 \times 10^{-2})$.
At $t \sim U/4$, the ratio $t^{\rm eff} /
\Delta^{\rm eff}$ begins to decreases and then
increases.
Therefore the conductance shows a broad peak near
the region $t \sim U/4$.
For $t \gsim U/4$ case, the effective parameters
behave $t^{\rm eff} \sim t$, $\Delta^{\rm eff}
\sim \Delta$.
We note that the conductance has the expression of
the non-interacting one itself in $t \gsim U/4$
region.

Finally we compare between the phase shift and the
occupation number shown in
Fig. \ref{fig:01}.
The phase shift of the odd orbital $\delta_{\rm
o}$ rapidly changes from $\pi/2$ to $0$ near $t
\sim 5 \times 10^{-4}$, even though $\langle
n_{\rm o} \rangle$ still remains at $\langle
n_{\rm o} \rangle \sim 1$.
Friedel's sum rule in each channel does not hold,
as already pointed out previously
\cite{rf:Sakai_2imp920102}.
It seems that this behavior is enhanced when the
anti-ferromagnetic coupling between the two sites
competes with the Kondo effect.

\subsection{From Weakly to Strongly correlated cases}
\label{sec:3.2}

In this subsection we present the numerical
results of the conductance for various $\Delta/\pi
U$ cases within $1.5 \times 10^{-2} \le \Delta/\pi
U \le 6.0 \times 10^{-2}$.
($1.7 \lsim u \lsim 6.8$.
The hybridization strength is changed in $1.5
\times 10^{-3} \le \Delta/\pi \le 6.0 \times
10^{-3}$, and the Coulomb repulsion is fixed at
$U=0.1$.)
We confirm the scenario shown in the previous
subsection that $J_{\rm LR}^{\rm eff} \sim T_{\rm
K}^{0}$ holds at the main peak of the conductance
with $t^{\rm eff} \sim \Delta^{\rm eff}$.
We also demonstrate how the kinetic exchange
process appear in the conductance for arbitrary
$\Delta/\pi U$ cases.

\begin{figure}[hbtp]
   \centerline{\epsfxsize=8.5cm\epsfbox{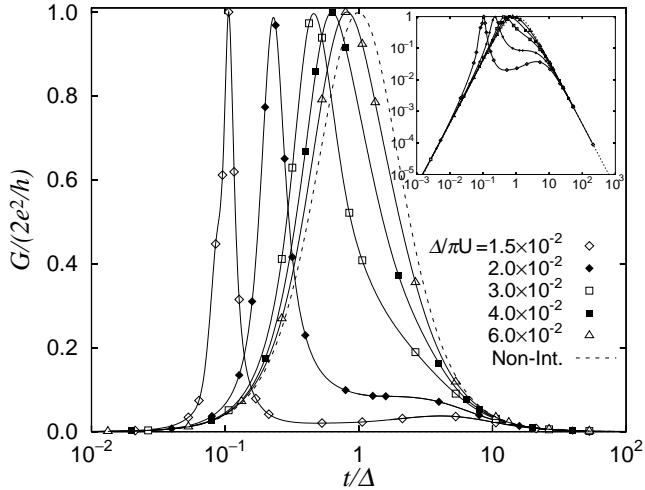}}
   \caption
   {
Conductance as a function of $t$ at zero
temperature, from the weakly to strongly
correlated cases in $1.5 \times 10^{-2} \le
\Delta/\pi U \le 6.0 \times 10^{-2}$.
The broken line shows the conductance for the
non-interacting (U=0) case.
   }
   \label{fig:04}
\end{figure}

The calculated conductance is shown in
Fig. \ref{fig:04}.
The horizontal axis is the hopping normalized by
the hybridization strength, $t/\Delta$.
From inset of the
Fig. \ref{fig:04}, the conductance
almost overlaps on the non-interacting curve in
the region $t \ll \Delta$ and $t \gg \Delta$.
Exactly, these regions should be classified
$J_{\rm LR}^{\rm eff} \ll T_{\rm K}^{0}$ and $t
\gg U/4$, respectively, from the analysis in the
previous subsection.
The conductance is very small in these regions,
however this uniform properties should be useful
to arrange the experimental data under uncertain
$U/\Delta$ cases.
%
%

All curves have a main peak with strength
$2e^{2}/h$.
For weakly correlated cases of $\Delta / \pi U
\gsim 4 \times 10^{-2}$, the conductance almost
coincides with the non-interacting one through all
$t$ region.
As $U/\Delta$ increases, the main peak shifts to
the small $t/\Delta$ side, and the peak width
becomes narrower.

We already found the relation $J_{\rm LR}^{\rm
eff} \sim T_{\rm K}^{0}$ at the main peak position
for $\Delta/\pi U =1.5 \times 10^{-2}$ case in \S
\ref{sec:3.1}.
Here we show the ratio $J_{\rm LR}^{\rm eff} /
T_{\rm K}^{0}$ at the main peak position for
various $\Delta/\pi U$ cases in Table
\ref{table:01}.
We can see the relation $J_{\rm LR}^{\rm eff}
\sim T_{\rm K}^{0}$ commonly \cite{rf:Izumida9901}.
(The relation at the main peak, $J_{\rm LR}^{\rm
eff} \sim T_{\rm K}^{0}$, would be generalized to
$E_{\rm B} \sim T_{\rm K}^{0}$ to the weakly
correlated cases, where $E_{\rm B} =
\sqrt{(2t)^{2} + (U/2)^{2}} - U/2$ is the singlet
binding energy between the two dots.)

From the analysis in the previous and present
subsection we can conclude the following effect of
the hopping term.
For the small $t$ case with $J_{\rm LR}^{\rm eff}
\ll T_{\rm K}^{0}$, ``(i) The Kondo singlet state
is formed on the left (right) dot with its
adjacent-site lead''.
On the other hand for the large $t$ case with
$J_{\rm LR}^{\rm eff} \gg T_{\rm K}^{0}$, ``(ii)
the local spins on each of the dots couple as the
singlet state''.
In the intermediate region, the Kondo effect of
the local spins on the orbitals extending on the
two dots (i.e. even and odd orbitals) occurs.
The main peak of the conductance appears around
the boundary between (i) and (ii) reflecting the
coherent connection of the leads and the dots.
As $U/\Delta$ increases, the Kondo temperature
$T_{\rm K}^{0}$ exponentially decreases, the
condition $J_{\rm LR}^{\rm eff} \sim T_{\rm
K}^{0}$ holds at the smaller $t/\Delta$, then the
main peak shifts to the smaller $t/\Delta$ side.
At the same time, the width of the peak becomes
extremely narrow compared with the decreasing of
$T_{\rm K}^{0}$.
This fact has been already shown as the steep
minimum of $\Delta^{\rm eff}$ in
Fig. \ref{fig:03}.
We note that this narrowing closely relates to the
quantum critical transition between the Kondo
singlet state and the local singlet state in the
two-impurity Kondo model \cite{rf:Jones8801}.
The shifting and narrowing behaviors shown here
are also pointed out with the SBMFT with
artificial addition of the anti-ferromagnetic
coupling between dots to the model
\cite{rf:Georges9901}.
However, the SBMFT calculation should be checked
by the method treating the kinetic exchange term
properly.
As noted in the introduction, the hopping term
causes two conflicting effects on the critical
transition of the two-impurity systems.
One is the kinetic exchange coupling $J_{\rm
LR}^{\rm eff}$, which causes the ``critical''
transition through the competition with the Kondo
effect.
Another is the parity splitting, which suppresses
the ``critical'' transition.
The calculation in this section is the first
reliable quantitative results of the two-impurity
Kondo problem in the DQD systems.

There is also another small peak (or shoulder)
structure for the strongly correlated cases of
$\Delta/\pi U \lsim 2 \times 10^{-2}$ ($u \gsim
5$) at larger $t$ side of the main peak.
In the previous subsection, we found that the
small peak appears around the boundary between
(ii) and (iii).
However for the weakly correlated cases, the small
peak could not be recognized because the condition
of the border (i)-(ii) and (ii)-(iii) could not be
distinguished clearly.

\end{multicols}
\widetext
\begin{table}[hbtp]
   \begin{center}
   \begin{tabular}{cccccc}
      $\Delta / \pi U$ & $1.5 \times 10^{-2}$ & $2 \times 10^{-2}$ & $3 \times 10^{-2}$ & $4 \times 10^{-2}$ & $6 \times 10^{-2}$ \\ \hline 
      $J_{\rm LR}^{\rm eff} / T_{\rm K}^{0}$ & $2.66$ & $2.34$ & $2.15$ & $2.09$ & $2.23$ \\
      $E_{\rm B} / T_{\rm K}^{0}$ & $2.66$ & $2.34$ & $2.14$ & $2.04$ & $2.05$
   \end{tabular}
   \end{center}
   \caption{
Ratios $J_{\rm LR}^{\rm eff} / T_{\rm K}^{0}$ and
$E_{\rm B} / T_{\rm K}^{0}$ at the main peak
position of the conductance.
}
   \label{table:01}
\end{table}
\begin{multicols}{2}
\narrowtext

\subsection{Temperature dependence of the conductance}
\label{sec:3.3}

In this subsection we present the conductance in
finite temperature.
We calculate the conductance at finite
temperatures by using the following formula
\cite{rf:Izumida9701};
\begin{eqnarray}
   G  & = &    \frac{2 e^{2}}{h} 
                \lim_{\omega \rightarrow 0} 
                \frac{P^{''}(\omega)}{\omega}. \label{eq:G_P''}
\end{eqnarray}
Here $P^{''}(\omega)$ is the `current spectrum'
for the current operator $J \equiv \dot{N_{\rm L}}
- \dot{N_{\rm R}}$ written as follows,
\begin{eqnarray}
   P^{''}(\omega) & = & 
                     \frac{\pi^{2} \hbar^{2}}{4}
                     \frac{1}{Z} \sum_{n, m} 
                     \left( {\rm e}^{- \beta E_{m}} - {\rm e}^{- \beta E_{n}} \right) 
   \nonumber \\
               && 
                     \times 
                     \left| \langle n \left| 
                     J                    
                     \right| m \rangle \right| ^{2} 
   \nonumber \\
               &&    \times 
                     \delta \left( \omega - \left( E_{n} - E_{m} \right) \right),
   \label{eq:P''}
\end{eqnarray}
where $\dot{N_{\rm L}}$ is the time
differentiation of the electron number in the left
lead, $Z = \sum_{n} e^{- \beta E_{n}}$ is the
partition function of the system, and $\beta$ is
the inverse of the temperature ($\beta=1/T$).

\begin{figure}[hbtp]
   \centerline{\epsfxsize=8.5cm\epsfbox{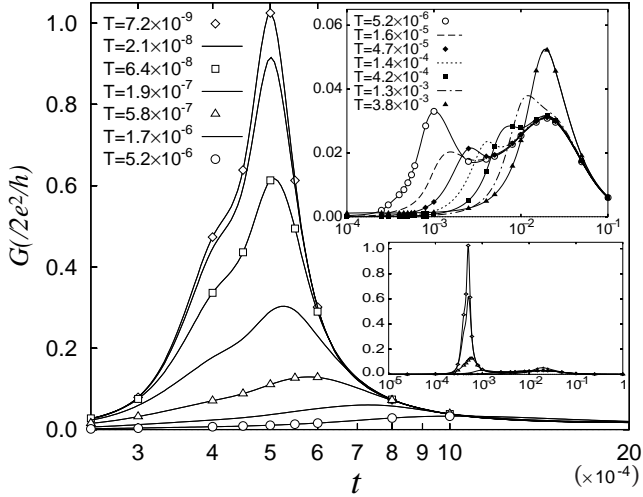}}
   \caption
   {
Temperature dependence of the conductance.
Main figure is the temperature dependence near the
main peak at $t \sim 5 \times 10^{-4}$.
Inset figure at the upper right is the temperature
dependence near the small peak at $t \sim 2.5
\times 10^{-2}$.
Inset figure at the lower right is the conductance
in all over $t$.
$\Delta/\pi U=1.5 \times 10^{-2}$.
   }
   \label{fig:05}
\end{figure}

First we show the conductance at various
temperatures for $\Delta/\pi U = 1.5 \times
10^{-2}$ case in
Fig. \ref{fig:05}.
As the temperature increases from $T \sim
10^{-8}$, the height of the main peak gradually
decreases.
At the same time the peak position shifts to the
larger $t$.
We note that $T \sim 10^{-8}$ is much lower than
$T_{\rm K}^{0}$.
($T_{\rm K}^{0} = 3.78 \times 10^{-6}$.)
%
%

\begin{figure}[hbtp]
   \centerline{\epsfxsize=8.5cm\epsfbox{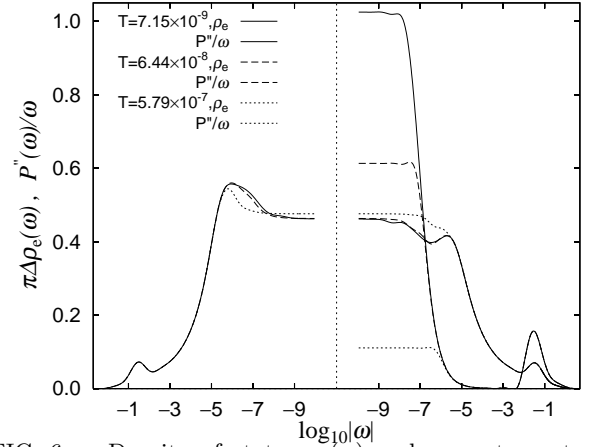}}
   \caption
   {
Density of states $\rho_{\rm e}(\omega)$ and
current spectrum $P^{''}(\omega)/\omega$ at $t =
5.0 \times 10^{-4}$ for finite temperatures.
(The spectrum, which has only the positive region
$\omega \ge 0$, is the current spectrum.)
   }
   \label{fig:06}
\end{figure}

\begin{figure}[hbtp]
   \centerline{\epsfxsize=8.5cm\epsfbox{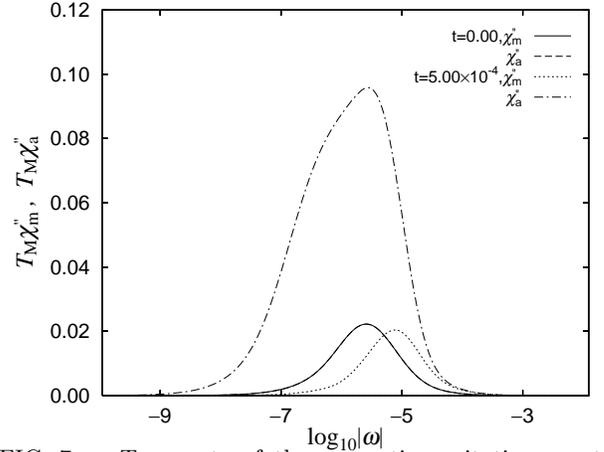}}
   \caption
   {
Two sorts of the magnetic excitation spectra
$\chi_{\rm m}^{''}(\omega)$ and $\chi_{\rm
a}^{''}(\omega)$ at $T=0$.
We note that the two spectra agree with each other
at $t=0$.
   }
   \label{fig:07}
\end{figure}

To discuss the characteristic behaviors of the
conductance in finite temperature, we show the
density of states $\rho_{\rm e}(\omega)$, and the
current spectrum (divided by $\omega$)
$P^{''}(\omega)/\omega$ at $t = 5.0 \times
10^{-4}$ in
Fig. \ref{fig:06}.
As the temperature increases to $T = 6.44 \times
10^{-8}$, $P^{''}(\omega)/\omega$ at $\omega \sim
0$ become $60 \%$ of $T=0$ limit.
At the same time, $\rho_{\rm e}(\omega)$ shows a
small change around $\omega \sim 10^{-7}$.
This means that the effect of the temperature on
the conductance is rather drastic.
Here we show two sorts of the magnetic excitation
spectra $\chi_{\rm m}^{''}(\omega)$ and $\chi_{\rm
a}^{''}(\omega)$ \cite{rf:Sakai_2imp920102},
where $\chi_{\rm m}^{''}(\omega)$ is the imaginary
part of the dynamical susceptibility of the
uniform magnetic moment of local spins, $(S_{{\rm
L}, z} + S_{{\rm R}, z})/\sqrt{2}$,
and $\chi_{\rm a}^{''}(\omega)$ is that of the
anti-ferromagnetic moment, $(S_{{\rm L}, z} -
S_{{\rm R}, z})/\sqrt{2}$, respectively.
We show the two magnetic excitation spectra at
$t=0$ and $t=5.0 \times 10^{-4}$ in
Fig. \ref{fig:07}.
At $t=0$, the two spectra agree with each other.
However at $t = 5.0 \times 10^{-4}$, $\chi_{\rm
a}^{''}(\omega)$ has the structure in lower energy
region than $\chi_{\rm m}^{''}(\omega)$.
It seems that $P^{''}(\omega)$ at the main peak of
the conductance is dominated by the fluctuation
given by $\chi_{\rm a}^{''}(\omega)$ from
Fig. \ref{fig:06}
and Fig. \ref{fig:07}.
%
%
%
%

We determine the two characteristic energies from
$\chi_{\rm m}^{''}(\omega)$ and $\chi_{\rm
a}^{''}(\omega)$ as the following ways.
One is determined from the peak position of
$\chi_{\rm m}^{''}(\omega)$, we call it $T_{\rm
M}$
\cite{rf:Izumida.SGL,rf:Izumida9801,rf:Izumida9701}.
Another one, we call it $T_{\rm AF}$, is
determined as $T_{\rm AF}/T_{\rm AF, 0} \equiv
X_{0}/X$, where $X \equiv \lim_{\omega \rightarrow
0} \chi_{\rm a}^{''}(\omega)/\omega$
\cite{rf:Sakai_2imp920102}.
(And here we have $T_{\rm AF, 0} \equiv T_{\rm M,
0}$.)
The suffix `$0$' indicate `$t=0$'.
The quantity $T_{\rm M, 0}$ almost coincides with
$T_{\rm K}^{0}$.
The ratio $T_{\rm M, 0}/T_{\rm K}^{0}$ for some
cases are shown in Ref. \cite{rf:Izumida.SGL}.

\begin{figure}[hbtp]
   \centerline{\epsfxsize=8.5cm\epsfbox{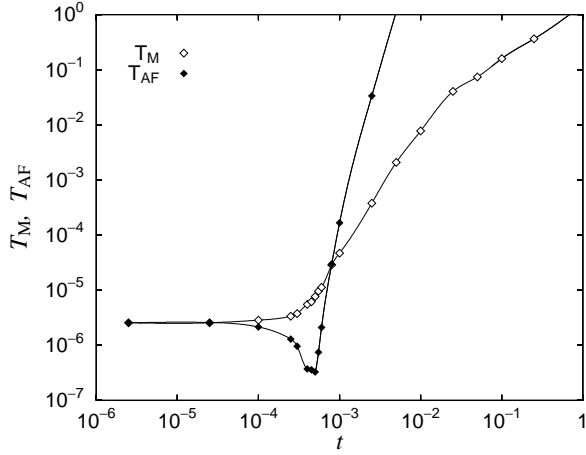}}
   \caption
   {
Two characteristic energies $T_{\rm M}$ and
$T_{\rm AF}$.
   }
   \label{fig:08}
\end{figure}

The calculated two characteristic temperatures,
$T_{\rm M}$ and $T_{\rm AF}$, are shown in
Fig. \ref{fig:08}.
They take almost same values in $t \lsim 10^{-4}$.
$T_{\rm M}$ monotonically increases as $t$
increasing.
On the other hand $T_{\rm AF}$ becomes smaller
near $J_{\rm LR}^{\rm eff} \sim T_{\rm K}^{0}$.
It has minimum of $T_{\rm AF} \simeq 3 \times
10^{-7}$ at $t \simeq 5 \times 10^{-4}$.
We note that the reduction of $T_{\rm AF}$ near
$T_{\rm K}^{0} \sim J$ had been already pointed
out \cite{rf:Sakai_2imp9001}.
As $t$ still increases, $T_{\rm AF}$ rapidly
increases.
From same analysis for the other $\Delta/\pi U$
cases, we confirm that the minimum of $T_{\rm AF}$
appears for the strongly correlated cases of
$\Delta/\pi U \lsim 2 \times 10^{-2}$.
We show $T_{\rm AF}$ at the main peak position in
Table. \ref{table:02}.
%
%

From the comparison with effective parameters in
Fig. \ref{fig:03},
the larger of the effective parameters,
$\max(t^{\rm eff}, \Delta^{\rm eff})$, and the
smaller of the characteristic temperature,
$\min(T_{\rm AF}, T_{\rm M})$, almost coincide
with each other in all $t$, $\max(t^{\rm eff},
\Delta^{\rm eff}) \sim \min(T_{\rm AF}, T_{\rm
M})$.

Here we again see the temperature dependence of
the conductance shown in
Fig. \ref{fig:05} with the
characteristic temperature shown in
Fig. \ref{fig:08}.
We can see that $T_{\rm AF}$ characterizes the
main peak of the conductance in finite
temperature.
The peak decreases as the temperature increases
near $T \sim 1 \times 10^{-8} (\sim 0.1 T_{\rm
AF}(t = 5 \times 10^{-4}) )$ in
Fig. \ref{fig:05}.
As the temperature increases and reaches to $T
\sim 1 \times 10^{-6} (\sim 10 T_{\rm AF}(t = 5
\times 10^{-4}) )$, the strength of the main peak
becomes almost zero.
Next we see the temperature dependence of the small
peak.
The small peak near $t \sim U/4 = 2.5 \times
10^{-2}$ increases as the temperature increases to
about $T \sim 10^{-3} (\sim 0.1 T_{\rm M}(t = 2.5
\times 10^{-2}) )$.
It seems that the characteristic temperature of
the conductance near the small peak is $T_{\rm
M}$.
From above it seems that the characteristic
temperature of the conductance is $\min(T_{\rm
AF}, T_{\rm M})$ in all $t$ region.

\begin{figure}[hbtp]
   \centerline{\epsfxsize=17cm\epsfbox{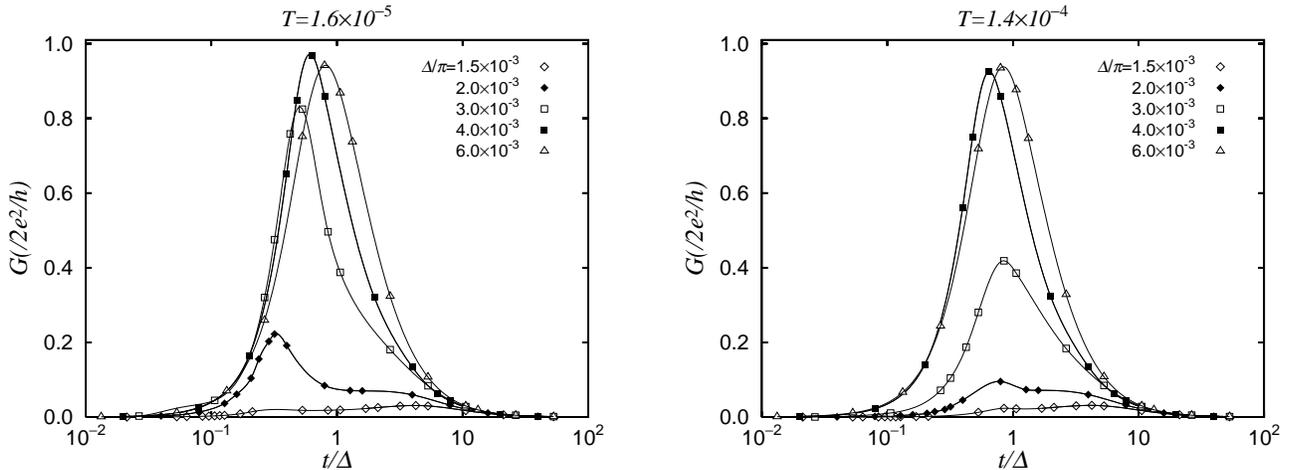}}
   \caption
   {
Conductance from weakly to strongly correlated
cases at $T=1.6 \times 10^{-5}$ (left figure) and
at $T=1.4 \times 10^{-4}$ (right figure).
   }
\label{fig:09}
\end{figure}

Here we show the conductance from the weakly to
strongly correlated cases at fixed temperatures.
We show the conductance at $T=1.6 \times 10^{-5}$
and $T=1.4 \times 10^{-4}$ in
Fig. \ref{fig:09}.
The main peak for the strongly correlated cases is
sensitive to the temperature.
Then the main peak of the conductance will shift
to smaller $t/\Delta$ side with increasing peak
height when the temperature decreases as seen from
Fig. \ref{fig:09}.
This behavior will be observed as the split gate
voltage is varied.
We note that $T = 1.4 \times 10^{-4}$ corresponds
to $16 {\rm mK}$, and $T=1.6 \times 10^{-5}$
corresponds to $1.9 {\rm mK}$, for $U = 1.0 {\rm
meV}$ systems.

\end{multicols}
\widetext
\begin{table}[hbtp]
   \begin{center}
   \begin{tabular}{cccccc}
      $\Delta / \pi U$ & $1.5 \times 10^{-2}$ & $2 \times 10^{-2}$ & $3 \times 10^{-2}$ & $4 \times 10^{-2}$ & $6 \times 10^{-2}$ \\ \hline 
      $T_{\rm AF}$ & $3.28 \times 10^{-7}$ & $9.77 \times 10^{-6}$ & $3.97 \times 10^{-4}$ & $1.50 \times 10^{2}$ & $3.01 \times 10^{2}$ \\
      $T_{\rm M}$ & $7.66 \times 10^{-6}$ & $5.64 \times 10^{-5}$ & $7.73 \times 10^{-4}$ & $3.47 \times 10^{-3}$ & $1.20 \times 10^{-2}$ \\
    ( $T_{\rm M, 0}$ & $2.55 \times 10^{-6}$ & $2.47 \times 10^{-5}$ & $2.31 \times 10^{-4}$ & $8.02 \times 10^{-4}$ & $3.22 \times 10^{-3}$ )
   \end{tabular}
   \end{center}
   \caption{
Characteristic energies at the main peak position.
}
   \label{table:02}
\end{table}
%
%
\begin{multicols}{2}
\narrowtext

\begin{figure}[hbtp]
   \centerline{\epsfxsize=8.5cm\epsfbox{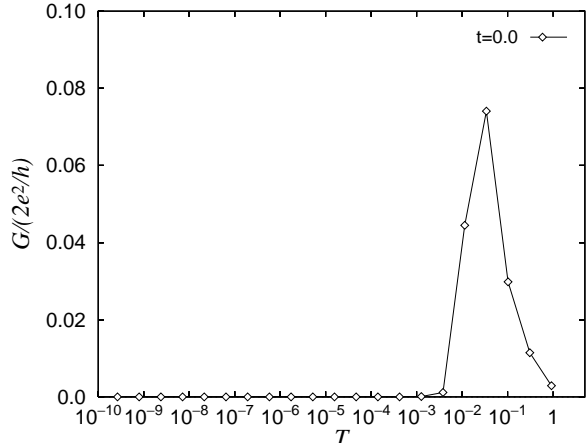}}
   \caption
   {
Numerical results of the conductance for $t = 0$
case.
   }
\label{fig:10}
\end{figure}

Finally, we note the accuracy of the conductance
calculated from
eqs. (\ref{eq:G_P''})-(\ref{eq:P''}) by using the
NRG method.
It is not so accurate at very high temperatures for
the small peak.
In the case of $t = 0$, two dots completely
decouple, then the conductance should be zero.
However as shows in
Fig. \ref{fig:10},
the calculated conductance has finite value in $5
\times 10^{-3} \lsim T \lsim 1$ and it has maximum
at $T \sim 0.05 (=U/2)$.
Thus the result at very high temperatures has
ambiguities.
This improper finite conductance would be caused
by estimation of eq. (\ref{eq:G_P''}) at $\omega
\sim T$ instead of $\omega \rightarrow 0$.
The finite value in the current spectrum at $T
\sim 0.05$ would reflect the largeness of the
dynamical charge fluctuation in the dots.
%
%

\section{Summary and Discussion}
\label{sec:4}

We calculated the tunneling conductance through
the two quantum dots that connected to the left
lead and the right lead in series.
We investigated the effect of the kinetic exchange
coupling between the dots, and also the
competition with the Kondo effect.

As the hopping between the two dots increased, (i)
Kondo singlet state, (ii) local spin singlet state,
and (iii) molecular orbital like state with double
occupancy in even state, continuously appeared.
For $t \ll U$ cases, the Kondo binding between the
left (right) lead and the left (right) dot,
$T_{\rm K}^{0}$, and the anti-ferromagnetic
kinetic exchange coupling between the two dots,
$J_{\rm LR}^{\rm eff}$, competed.
The boundary between (i) and (ii) was characterized
as $J_{\rm LR}^{\rm eff} \sim T_{\rm K}^{0}$, and
the tunneling conductance showed a peak.
This peak had the unitarity limit value of
$2e^{2}/h$ reflecting the coherent connection
through the lead-dot-dot-lead.
At $t \sim U/4$ of the boundary between (ii) and
(iii), we had a small peak.

The system showed the strongly correlated
behaviors for $\Delta/\pi U \lsim 2 \times
10^{-2}$ ($u \equiv U/\pi \Delta \gsim 5$) cases.
The borders of (i)-(ii) ($J_{\rm LR}^{\rm eff}
\sim T_{\rm K}^{0}$) and (ii)-(iii) ($t \sim U/4$)
were clearly distinguished, then there were the
two peak structures in the conductance.
Furthermore the width of the main peak became
steeply narrow.
The characteristic temperature of the main peak
was strongly reduced compared with the Kondo
temperature of the single dot systems $T_{\rm
K}^{0}$.
These anomalous behaviors of the main peak related
to the quantum critical transition of the
two-impurity Kondo problem studied in previously.
Though the hopping term had conflicting effects on
the critical transition of the two-impurity Kondo
systems, generation of it through the kinetic
exchange coupling and suppression of it due to the
parity splitting, we found that we see the sign of
the anomaly in the tunneling conductance.

The quantitative calculation shown in this paper
gave the new realization for the two-impurity
Kondo problem.
This investigation suggested the importance of the
systematic study of the DQD systems for the
two-impurity Kondo problem.

\section*{Acknowledgments}

This work was partly supported by a Grant-in-Aids
for Scientific Research on the Priority Area
``Spin Controlled Semiconductor Nanostructures''
(No. 11125201) from the Ministry of Education,
Science, Sports and Culture.
The numerical computation was partly performed at
the Supercomputer Center of Institute for Solid
State (University of Tokyo), the Computer Center
of Institute for Molecular Science (Okazaki
National Research Institute), and the Computer
Center of Tohoku University.
One of the author (W. I.) is supported by JSPS
fellowship.


\end{multicols}


\begin{thebibliography}{99}

\bibitem[*]{email}
izumida@eng.hokudai.ac.jp

\bibitem{rf:Hewson}
A. C. Hewson: {\it The Kondo Problem to Heavy Fermions}
(Cambridge University Press, 1993).

\bibitem{rf:Yosida}
K. Yosida: {\it Theory of Magnetism} (Springer,
1996).


\bibitem{rf:Jones8701}
B. A. Jones, C. M. Varma and J. W. Wilkins:
Phys. Rev. Lett. {\bf 58}, 843 (1987).

\bibitem{rf:Jones8801}
B. A. Jones, C. M. Varma and J. W. Wilkins:
Phys. Rev. Lett. {\bf 61}, 125 (1988).

\bibitem{rf:Jones8901}
B. A. Jones and C. M. Varma: Phys. Rev. B {\bf
40}, 324 (1989).

\bibitem{rf:Jones9101}
B. A. Jones: Physica B {\bf 171}, 53 (1991).

\bibitem{rf:Jones8902}
B. A. Jones, B. G. Kotilar and A. J. Mills:
Phys. Rev. B {\bf 39}, 3415 (1989).


\bibitem{rf:Sakai_2imp9001}
O. Sakai, Y. Shimizu and T. Kasuya: Solid State
Commun. {\bf 75}, 81 (1990).

\bibitem{rf:Sakai_2imp920102}
O. Sakai and Y. Shimizu: J. Phys. Soc. Jpn. {\bf
61}, 2333 (1992); {\bf 61}, 2348 (1992).


\bibitem{rf:Fye8701}
R. M. Fye, J. E. Hirsch and D. J. Scalapino: Phys. 
Rev. B {\bf 35}, 4901 (1987).

\bibitem{rf:Fye8901}
R. M. Fye and J. E. Hirsch: Phys. Rev. B {\bf 40},
4780 (1989).

\bibitem{rf:Fye9401}
R. M. Fye: Phys. Rev. Lett. {\bf 72}, 916 (1994).


\bibitem{rf:Affleck9201} 
I. Affleck and A. W. W. Ludwig:
Phys. Rev. Lett. {\bf 68}, 1046 (1992).

\bibitem{rf:Affleck9501} 
I. Affleck, A. W. W. Ludwig and B. A. Jones: Phys. 
Rev. B {\bf 52}, 9528 (1995).







\bibitem{rf:DGG9801}
D. Goldhaber-Gordon, H. Shtrikman, D. Mahalu,
D. Abusch-Magder, U. Meirav and M. A. Kastner:
Nature {\bf 391}, 156 (1998).

\bibitem{rf:Cronenwett9801}
Sara M. Cronenwett, Tjerk H. Oosterkamp and Leo P. 
Kouwenhoven: Science {\bf 281}, 540 (1998).

\bibitem{DGG9802}
D. Goldhaber-Gordon, J. G{$\ddot{\rm o}$}res,
M. A. Kastner, H. Shtrikman, D. Mahalu and
U. Meirav: Phys. Rev. Lett.  {\bf 81}, 5225
(1998).

\bibitem{rf:Schmid9801}
J. Schmid, J. Weis, K. Eberl and K. Klitzing:
Physica B {\bf 256-258}, 182 (1998).

\bibitem{rf:Simmel9901}
F. Simmel, R. H. Blick, J. P. Kotthaus,
W. Wegscheider and M. Bichler: Phys. Rev. Lett.
{\bf 83}, 804 (1999).


\bibitem{rf:Izumida.SGL} 
O. Sakai, W. Izumida and S. Suzuki: Proceedings of
the 4-th Int. Symposium on Advanced Physical Field
(March, 1999, Tsukuba, Japan), 143;
W. Izumida and O. Sakai: Physica B (in press,
Proceedings of the International Conference on
Strongly Correlated Electron Systems, (August,
1999, Nagano, Japan));
W. Izumida, O. Sakai and S. Suzuki: in
preparation.


\bibitem{rf:EXP_2DOT}
For example of the recent experimental studies of
the double dot systems, T. H. Oosterkamp,
T. Fujisawa, W. G. van der Wiel, K. Ishibashi,
R. V. Hijman, S. Tarucha and L. P. Kouwenhoven:
Nature {\bf 395}, 873 (1998);
T. Fujisawa, T. H. Oosterkamp, W. G. van der Wiel,
B. W. Broer, R. Aguado, S. Tarucha and
L. P. Kouwenhoven: Science {\bf 282}, 932 (1998).


\bibitem{rf:Ivanov9701}
T. Ivanov: Europhys. Lett. {\bf 40}, 183 (1997).

\bibitem{rf:Pohjola9701}
T. Pohjola, J. K{\"o}nig, M. M. Salomaa,
J. Schmid, H. Schoeller and Gerd. Sch{\"o}n:
Europhys. Lett. {\bf 40}, 189 (1997).

\bibitem{rf:Aono9801}
T. Aono, M. Eto and K. Kawamura: J. Phys. Soc.
Jpn. {\bf 67}, 1860 (1998).

\bibitem{rf:Izumida9901} W. Izumida, O. Sakai and Y.
Shimizu: Physica B {\bf 259-261}, 215 (1999).

\bibitem{rf:Georges9901}
A. Georges and Y. Meir: Phys. Rev. Lett. {\bf 82},
3508 (1999).

\bibitem{rf:Izumida.LT22}
W. Izumida and O. Sakai: Physica B (in press,
Proceedings of the XXII International Conference
on Low Temperature Physics, (August, 1999, Espoo
and Helsinki, Finland)).


\bibitem{rf:Izumida9801} 
W. Izumida, O. Sakai and Y. Shimizu: J. Phys. Soc. 
Jpn. {\bf 67}, 2444 (1998).

\bibitem{rf:Izumida9701}
W. Izumida, O. Sakai and Y. Shimizu: J. Phys. Soc. 
Jpn. {\bf 66}, 717 (1997).


\end{thebibliography}
\end{document}